\documentclass{article}

\usepackage{arxiv}
\usepackage[utf8]{inputenc} % allow utf-8 input
\usepackage[T1]{fontenc}    % use 8-bit T1 fonts
\usepackage{hyperref}       % hyperlinks
\usepackage{url}            % simple URL typesetting
\usepackage{booktabs}       % professional-quality tables
\usepackage{amsfonts}       % blackboard math symbols
\usepackage{nicefrac}       % compact symbols for 1/2, etc.
\usepackage{microtype}      % microtypography
\usepackage{graphicx}
\usepackage{doi}
\usepackage{amsmath}
\usepackage{natbib}
\usepackage{url}

\DeclareUnicodeCharacter{0394}{\ensuremath{\Delta}}

\title{A coupled model for the linked dynamics of marine pollution by microplastics and plastic-related organic pollutants}

\author{ \href{https://orcid.org/0000-0003-1449-1587}{\includegraphics[scale=0.06]{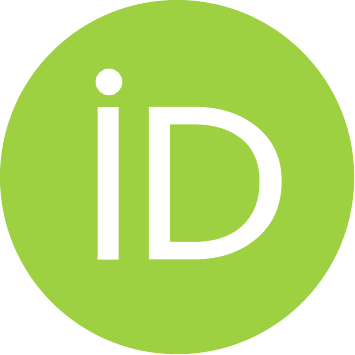}\hspace{1mm}Federica Guerrini}\thanks{Corresponding authors.} \\
	Dipartimento di Elettronica, Informazione e Bioingegneria\\
	Politecnico di Milano\\
	Via Ponzio 34/5, 20131 Milan (Italy) \\
	\texttt{federica.guerrini@polimi.it} \\
	\And
	\href{https://orcid.org/0000-0003-1326-9992}{\includegraphics[scale=0.06]{orcid.pdf}\hspace{1mm}Lorenzo Mari} \\
	Dipartimento di Elettronica, Informazione e Bioingegneria\\
	Politecnico di Milano\\
	Via Ponzio 34/5, 20131 Milan (Italy) \\
	\texttt{lorenzo.mari@polimi.it} \\
	\AND
		\href{https://orcid.org/0000-0001-5177-803X}{\includegraphics[scale=0.06]{orcid.pdf}\hspace{1mm}Renato Casagrandi*} \\
	Dipartimento di Elettronica, Informazione e Bioingegneria\\
	Politecnico di Milano\\
	Via Ponzio 34/5, 20131 Milan (Italy) \\
	\texttt{renato.casagrandi@polimi.it} \\
}

\date{}

\hypersetup{
pdftitle={TA coupled model for the linked dynamics of marine pollution by microplastics and plastic-related organic pollutants},
pdfsubject={Lagrangian-Eulerian model},
pdfauthor={Federica Guerrini, Lorenzo Mari, Renato Casagrandi},
pdfkeywords={microplastics, marine pollution, Lagrangian modelling, Eulerian modeling, advection-diffusion,hydrophobic pollutants},
}

\begin{document}
\maketitle

\begin{abstract}
	The pervasiveness of microplastics in global oceans is raising concern about its impacts on organisms. While quantifying its toxicity is still an open issue, sampling evidence has shown that rarely is marine microplastics found clean; rather, it is often contaminated by other types of chemical pollutants, some known to be harmful to biota and humans. To provide a first tool for assessing the role of microplastics as vectors of plastic-related organic pollutants (PROPs), we developed a data-informed model that accounts for the intertwined dynamics of Lagrangian microplastic particles transported by surface currents and the Eulerian advection-diffusion of chemicals that partition on them through seawater-particle interaction. Focusing on the Mediterranean Sea and using simple, yet realistic forcings for the input of PROPs, our simulations highlight that microplastics can mediate PROP export across different sub-seas. Particle origin, in terms of both source type (either coastal, riverine, or fishing-derived) and geographical location, seems to play a major role in determining the amount of PROPs conveyed by microplastics during their journey at sea. We argue that quantitative numerical modelling approaches can be focal to shed some light on the vast spatial and temporal scales of microplastics-PROPs interaction, complementary to much-needed on-field investigation. 
\end{abstract}

\keywords{microplastics \and marine pollution \and Lagrangian modelling \and Eulerian modeling \and advection-diffusion \and hydrophobic pollutants}

\section{Introduction}
Microplastics (plastic fragments smaller than 5~mm, \citealp{Arthur2008}) are emerging as a new class of marine contaminants \citep{Rochman2019}. Not only are they composed of various types of polymers and chemical additives, as diversified as the products they originated from, they also are altered during their permanence in the environment. Weathering affects microplastics' size and morphology \citep{Andrady2011,Kalogerakis2017,Song2017,Efimova2018}, and can lead to biological interactions, such as biofouling by bacteria and algae \citep[sometimes including invasive species and pathogens,][]{Barnes2002,Zettler2013} and incorporation into aggregates \citep{Long2015,DeHaan2019,Kvale2020a}. Focal to this study, the chemical profiles of microplastics may also change during their journey at sea, since microplastics have shown to leach plastic additives and have high sorptive capacities for environmental contaminants \citep{Mato2001,Teuten2007,Engler2012,Velzeboer2014,Worm2017,Galloway2017}. The mounting, evidence-based awareness of microplastics' relevance as vectors of contaminants is raising concern \citep{Ziccardi2016,Hartmann2017,Leon2018,Leon2019,Menendez-Pedriza2020}, as such microscale interactions may determine toxicity to organisms scaling up to macroscale (e.g.~\citealp{Werner2016}). The mechanical risks caused by microplastic particles to marine biota, such as blockage and/or reduced nutrition \citep{Wright2013,Cole2015,Galloway2016,Ogonowski2016}, up to translocation in tissues (see e.g.~\citealp{Browne2007,Mattsson2017}, and in humans, \citealp{Ragusa2021}), may be exacerbated by chemical risk. Not only can microplastics function, on top of natural pathways, as an additional exposure medium to environmental pollutants, they can also have synergistic toxic effects on organisms when combined with harmful chemicals---two points of concern still at the frontier of relevant research \citep{Teuten2009,Fossi2012,Koelmans2013,Koelmans2016,Diepens2018,Gallo2018,MohamedNor2019,Tetu2019,Menendez-Pedriza2020}.

Field knowledge of marine microplastic pollution has often taken insights from numerical models \citep{Hardesty2017}. Particle-tracking models at oceanographic scales are widely used to describe microplastic transport and distribution on the sea surface \citep[e.g.~][]{Lebreton2012,Eriksen2014,VanSebille2015,Liubartseva2018,Soto-Navarro2020,Onink2021,Guerrini2021} and along the water column, also accounting for biological interactions \citep{Kooi2017,Lobelle2021}. Besides these physical aspects, none of these models has yet been used to assess the chemical interactions occurring during the journey of plastic fragments at sea. These phenomena add a new dimension of complexity to the modelling of microplastic pollution: accounting for environmentally relevant changes caused by chemical-physical interactions to both microplastics and seawater during the journey of each fragment. Here we propose a novel modelling framework for microplastic pollution based on the coupling of particle tracking, transport of plastic-related organic pollutants (PROPs), and microplastic-seawater chemical exchanges, to describe their linked spatiotemporal dynamics at a basin-wide scale. In so doing, we identify areas where the exchange of PROPs operated by microplastics modulates seawater pollution. Thanks to our particle-based approach, these contributions can be apportioned to their country and source-type of origin (coastal, riverine, or fishing-related). Additionally, the trajectories of the top-adsorbing or -desorbing particles are analyzed to detect recurrent spatial patterns possibly driving microplastics-PROPs interactions at a basin scale.

\section{Methods}
\label{sec:2}

\begin{figure}
	\centering
	\includegraphics[width=90mm]{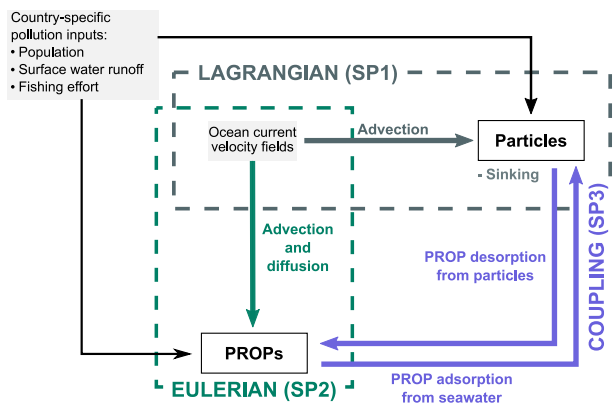}
	\caption{Diagram of our coupled Lagrangian-Eulerian framework. Model variables, i.e.~plastic particles and PROPs, are framed in black. Thick color-coded arrows represent sub-problems (grey: SP1, green: SP2, lilac: SP3). Thin arrows represent data used to force the model (in black).}
	\label{fig:1}
\end{figure}

We identified three separate sub-problems (SPs) to be solved when considering the coupled dynamics of microplastic particles and PROPs in marine environments (Fig.~\ref{fig:1}): 
\begin{itemize}
\item SP1 -- particle transport by sea surface currents, 
\item SP2 -- advection-diffusion of PROPs on the sea surface, and 
\item SP3 -- gradient-driven mass transfer of PROPs between particles and seawater.
\end{itemize} 
Numerical simulations of our coupled algorithm were designed using an operator-splitting approach \citep[{\it sensu} ][]{Strang1968}, whereby each SP was solved over staggered time steps. More precisely, the temporal sequence of subproblems execution in our algorithm is as follows:
\begin{enumerate}
\item Eulerian advection-diffusion of PROPs over the oceanographic domain (SP2) is solved over a half time-step $[0;\frac{\Delta t}{2}]$; 
\item Lagrangian particle transport by surface currents (SP1) is simulated at a Mediterranean-wide scale over an entire time step  $[0;\Delta t]$; 
\item Coherently with the chemical gradient between PROP concentration on each particle and in the seawater surrounding it at time $t=\Delta t$, water-particle pollutant exchanges are then solved at particle-scale in $[0;\Delta t]$ and PROP concentration in seawater is updated (SP3);
\item SP2 is solved again for another half time step $[\frac{\Delta t}{2};\Delta t]$, so that the numerical scheme is completed over a full time step.
\end{enumerate}
This algorithm is then executed throughout the duration of the simulation.

\subsection*{SP1: Lagrangian particle tracking simulations}
\label{sec:SP1}
 We exploited the data-informed Lagrangian model presented in \citep{Guerrini2021}, briefly recalled in the following (see also the Supplementary Materials). As commonly done, particles were treated in SP1 as point-like entities that move coherently with marine currents. Drivers of particle movement were the velocity vectors obtained at particle coordinates by interpolating with a tri-linear scheme the zonal and meridional components of the daily surface currents from the Mediterranean Sea Physical Reanalysis (1/16$^{\circ}$~x~1/16$^{\circ}$ horizontal resolution and 72 layers of depth; \citealp{Simoncelli2014}) available through Copernicus Marine Environment Monitoring Service. New particles were released daily from significant sources of marine plastic pollution, of either land-based origin, as coastlines and rivers \citep{Jambeck2015,Lebreton2017,IUCN2020}, or offshore origin, like fishing grounds \citep{Conservancy2011}. We adopted data-informed, source-specific proxies to modulate the spatiotemporal release of around 100 million particles per year from these sources (SM, Section~S1). The vertical position of each particle was assigned randomly upon release in the uppermost layer of the reanalyses (1.5--4.6~meter deep) and was then kept constant throughout the simulation, without accounting for daily vertical velocities. The key vertical movement accounted for is sinking: microplastic particles are eventually removed from the sea surface because of a decrease in buoyancy due to the growth of biofilm \citep{Fazey2016,Kooi2017,Amaral-Zettler2020} and/or aggregation in marine snow \citep{Cole2013,Andrady2015,Long2015,Michels2018,Kvale2020a}. The sinking of microplastics was simulated by attributing to each particle a transport duration randomly extracted from an exponential distribution with an average of 50 days \citep{Liubartseva2018}, and not exceeding 250 days. This resulted in a median sinking time of 34 days, a value coherent with the results by \cite{Fazey2016} and \cite{Lobelle2021}. Another mechanism acting on the balance of microplastics is beaching \citep{Eunomia2016,Kaandorp2020}. In our model, a particle was considered beached (and its simulation stopped) if it reached a position where both the zonal and meridional components of the velocity field were null. Re-suspension of beached particles (e.g.~because of rough sea or tidal motion) was not considered. It is worth remarking that particles' motion do not depend on chemical exchanges. Thus, SP1 could be run first and with no reference to the other SPs. 

\subsection*{SP2: Eulerian advection-diffusion} 
\label{sec:SP2} 
To model the dispersion of PROPs within the surface layer of the Mediterranean Sea, we used a simple Eulerian approach based on the general advection-diffusion equation \citep{Fischer1979}
\begin{equation}
\label{eq:1}
\frac{\partial c}{\partial t} = \nabla(D\nabla c) - \nabla(\mathbf{v}c) + S
\end{equation}
where $c$ represents PROP concentration, $D$ is the diffusivity coefficient, $\textbf{v}$ is the velocity vector, and $S$ is a location-dependent input term (SM, Section~S1). To numerically solve Eq.~\ref{eq:1}, we used the cell-centered finite volume method implemented in the PDE solver FiPy (\href{https://www.ctcms.nist.gov/fipy/}{https://www.ctcms.nist.gov/fipy/}, \citealp{Guyer2009}). For spatial coherence with SP1, we defined a control volume with the same grid size (1/16$^{\circ}$ x 1/16$^{\circ}$) and depth (3.1 m) used there. At present, data scarcity prevented us from informing PROP inputs with field observations of hydrophobic pollutants at a whole-Mediterranean scale \citep{Guerrini2021}. To parameterize the source term of Eq.~\ref{eq:1}, in SP2 we released a target PROP (phenanthrene) daily from the same sources as in SP1, following the proxies that guided the release of microplastics. Despite its simplicity, our Eulerian module proved to yield realistic PROP concentrations on the sea surface \citep[][ and SM, Section~S2]{Guerrini2021}. Contrary to what remarked above for SP1, the pollutant exchanges occurring at the particle scale (see SP3 below) do affect PROP concentrations in seawater, thus they interact with the advection-diffusion processes occurring at basin scale (SP2). Therefore, SP2 and SP3, both informed by SP1, had to be executed simultaneously.

\subsection*{SP3: Modelling particle-seawater mass transfer}
\label{sec:SP3}
To quantify the mass transfer of pollutants between microplastics and seawater, particles transported in SP1 were qualified in SP3 with fixed properties (like polymer type, prototypical size, and shape of microplastics; \citealp{Rochman2015}) to (i) allow a realistic microplastic-seawater PROP partitioning, and (ii) more accurately describe transfer kinetics (SM, Section~S3). Under our assumptions, PROP exchange can be modelled by applying Fick's law to each microplastic particle, i.e.
\begin{equation}
\label{eq:2}
    \frac{d C_{i,j}(t)}{d t} = k_{des}(K_{P-W}C_{W,j}(t)-C_{i,j}(t))
\end{equation}
where ${C_{i,j}}$ represents PROP concentration on particle $i$ within cell $j$, $C_{W,j}(t)$ is the PROP concentration in the seawater of cell $j$ as computed in SP2. $K_{P-W}$ and $k_{des}$ are the plastic-water partition coefficient and the desorption rate, respectively; both depend on PROP-polymer pairing (SM, Section~S3.2). We use synchronous updating of $C_{i,j}(t)$ for all particles in the cell while keeping frozen $C_{W,j}(t)$, an assumption made plausible by the widely different spatial scales at which SP2 and SP3 occur. Eq.\ref{eq:2} is solved numerically using the Python package Scipy (\href{https://docs.scipy.org/doc/}{https://docs.scipy.org/doc}) with an explicit Runge-Kutta method of order 5(4). By summing up all the mass of PROPs exchanged (i.e.~adsorbed or desorbed) between particles and seawater in cell~$j$, it is possible to update the environmental concentration of PROPs (SM, Section~S3), which concludes the solution of SP3.

\subsection*{Numerical experiments}
By intertwining SP1, SP2, and SP3 at daily temporal resolution, we simulated two years (2015--2016) of coupled Lagrangian-Eulerian interactions between particles transported by Mediterranean surface currents and a target PROP undergoing advection-diffusion on the sea surface layer. To set credible, basin-wide initial conditions for the coupled simulation, SP1 and SP2 required a warm-up. As for SP1, we began the Lagrangian simulation 250 days before January 1$^{st}$, 2015, starting from a sea surface clean of microplastic particles. This duration corresponds to the longest residence time of particles floating within the sea surface layer, after which the Lagrangian part of the model is effectively memoryless. All particles were assumed to be uncontaminated at release. Since the advection-diffusion of PROPs is not memoryless, we used a longer warm-up period (2005--2014) for SP2, starting with a clean Mediterranean Sea and daily releasing a consistent quantity of our PROP from each source type and location (SM, Section~S2). During our coupled Lagrangian-Eulerian experiment, we overall tracked trajectories, particle-seawater PROP fluxes, and daily PROP concentrations of more than 250 million microplastic particles.

\section{Results}
\label{sec:3}
\begin{figure}
	\centering
	\includegraphics[width=90mm]{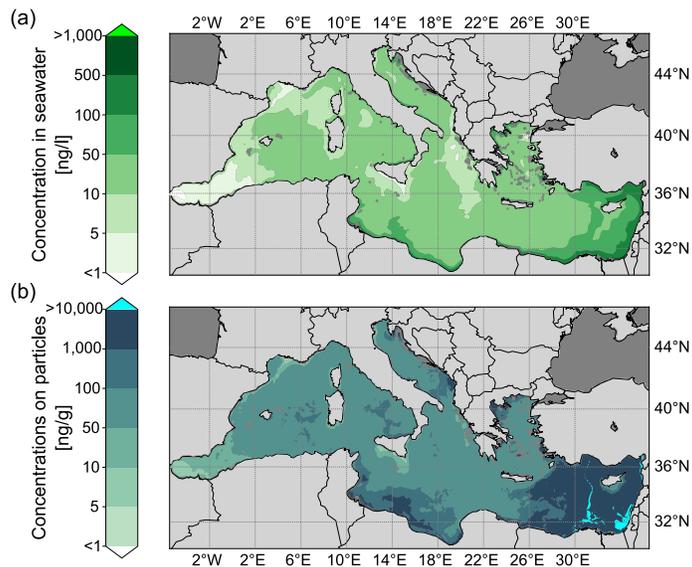}
	\caption{Concentrations of a target PROP (a) in seawater, or (b) onto particles as resulting from the coupled Lagrangian-Eulerian simulation, temporally averaged over 2015–2016.}
	\label{fig:2}
\end{figure}

Maps of the concentration (2015-–2016 average) of the target PROP, as simulated by our coupled Lagrangian-Eulerian model, are shown in Fig.~\ref{fig:2}. PROP concentration in surface seawater (\ref{fig:2}(a)) follows a rather smooth, west-to-east increasing gradient. The most notable local exceptions to that trend are the Gulf of Lyon, the Albanian and Ionian Greek coasts, and the southern Sicilian shoreline. Coastal areas tend to have higher PROP concentrations than off-coast. A west-to-east increase of pollution emerges also for particle-bound PROP concentration (\ref{fig:2}(b)). Some remarkable differences with \ref{fig:2}(a) exist, the most evident being the Levantine basin, where the modelled particle-bound PROP concentrations typically exceed 1,000 ng/g. Exceptional values of PROP adsorbed onto particles (larger than 10,000 ng/g) are all located in the Levantine area, sometimes showing strong geographic signatures (e.g.~the cyan path-like pattern from the Nile Delta to the southern Turkish coast). 

\begin{figure}
	\centering
	\includegraphics[width=180mm]{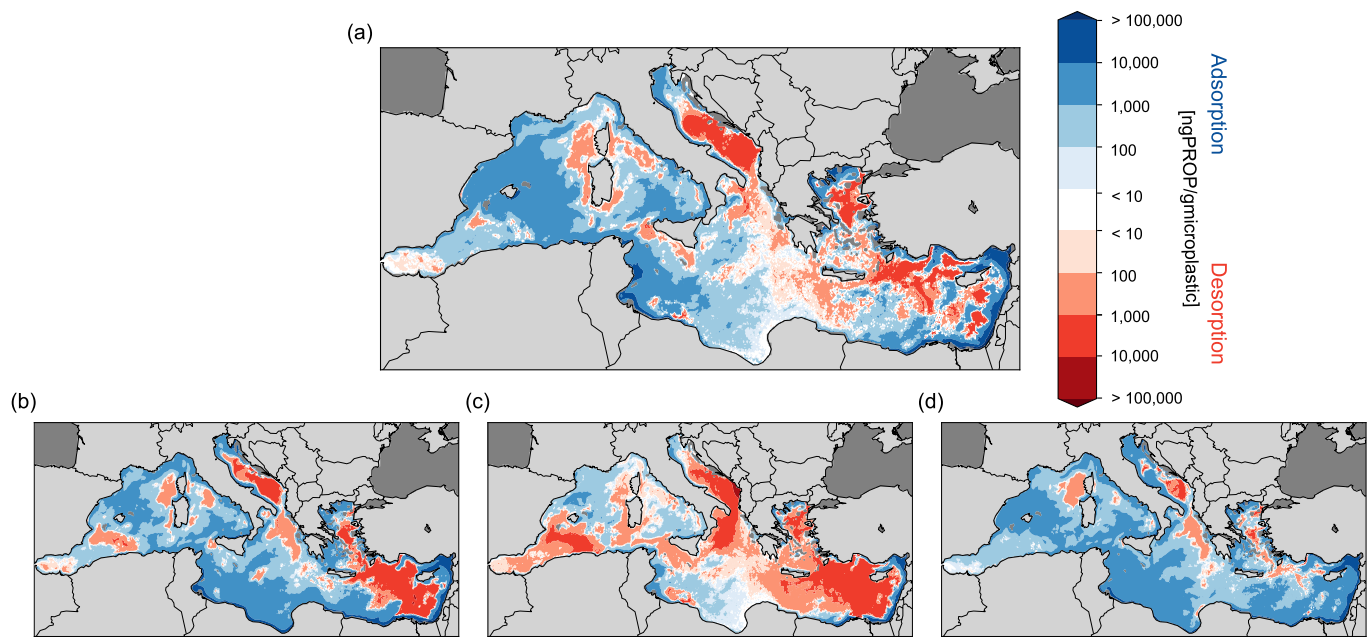}
	\caption{Net particle-mediated PROP exchanges averaged over 2015-–2016 as resulting from the coupled simulation. Blue shades indicate cells where adsorption of seawater PROP onto particles dominates over desorption from them (viceversa for red shades). Exchanges mediated by particles released from (a) all sources, (b) coasts, (c) rivers, and (d) fishing activities.}
	\label{fig:3}
\end{figure}

Our coupled model also allows to map the chemical exchanges between seawater and particles (Fig.~\ref{fig:3}). Net adsorption/desorption of the focal PROP onto/from particles, integrated over the entire simulation, are distributed in vast, spatially clustered areas in each regional sea within the Mediterranean basin. High values of net adsorption per gram of microplastics are typically recorded along the coasts, especially in the western Mediterranean (\ref{fig:3}(a)). By contrast, there are large regions in the central and eastern Mediterranean Sea where particles tend to release the adsorbed PROP, like in the south-central part of the Adriatic Sea, the Aegean Sea, and the northern Levantine Sea. Water-particle PROP fluxes appear to be weaker in the wide portion of the central Mediterranean (from the Ionian Sea to the east of Libya) than anywhere else. Although particles from all source types were clean of the PROP upon release, patterns of net adsorption/desorption differ when particles are partitioned by their source type. Riverine particles (\ref{fig:3}(c)) appear to be responsible for important releases of PROP throughout the Mediterranean basin, contributing both to areas characterized by high desorption and to areas with weaker overall desorption, like in the south of the Balearic islands. Coastal particles (\ref{fig:3}(b)) also show high desorption in the Adriatic and Levantine seas, yet in this case PROP adsorption prevails, especially in the western and southern parts of the Mediterranean basin. Particles from fishing vessels (\ref{fig:3}(d)) seem to mostly adsorb the simulated PROP, a result that is not surprising since fishing-related particles enter (clean, like all others) the sea from offshore locations, which are generally less polluted than coastal waters (see Fig.~\ref{fig:2}(a)). For this reason, and because of slower kinetics due to a smaller gradient between the particle-bound and seawater PROP concentration (see Eq.~\ref{eq:2}), particles released in fishing grounds are more likely to be chemically downgradient (i.e.~they tend to adsorb) during their transport at sea. However, some areas with significant values of specific desorption can be identified in \ref{fig:3}(d), mostly in the same subseas as in \ref{fig:3}(b) and \ref{fig:3}(c).

\begin{figure}
	\centering
	\includegraphics[width=180mm]{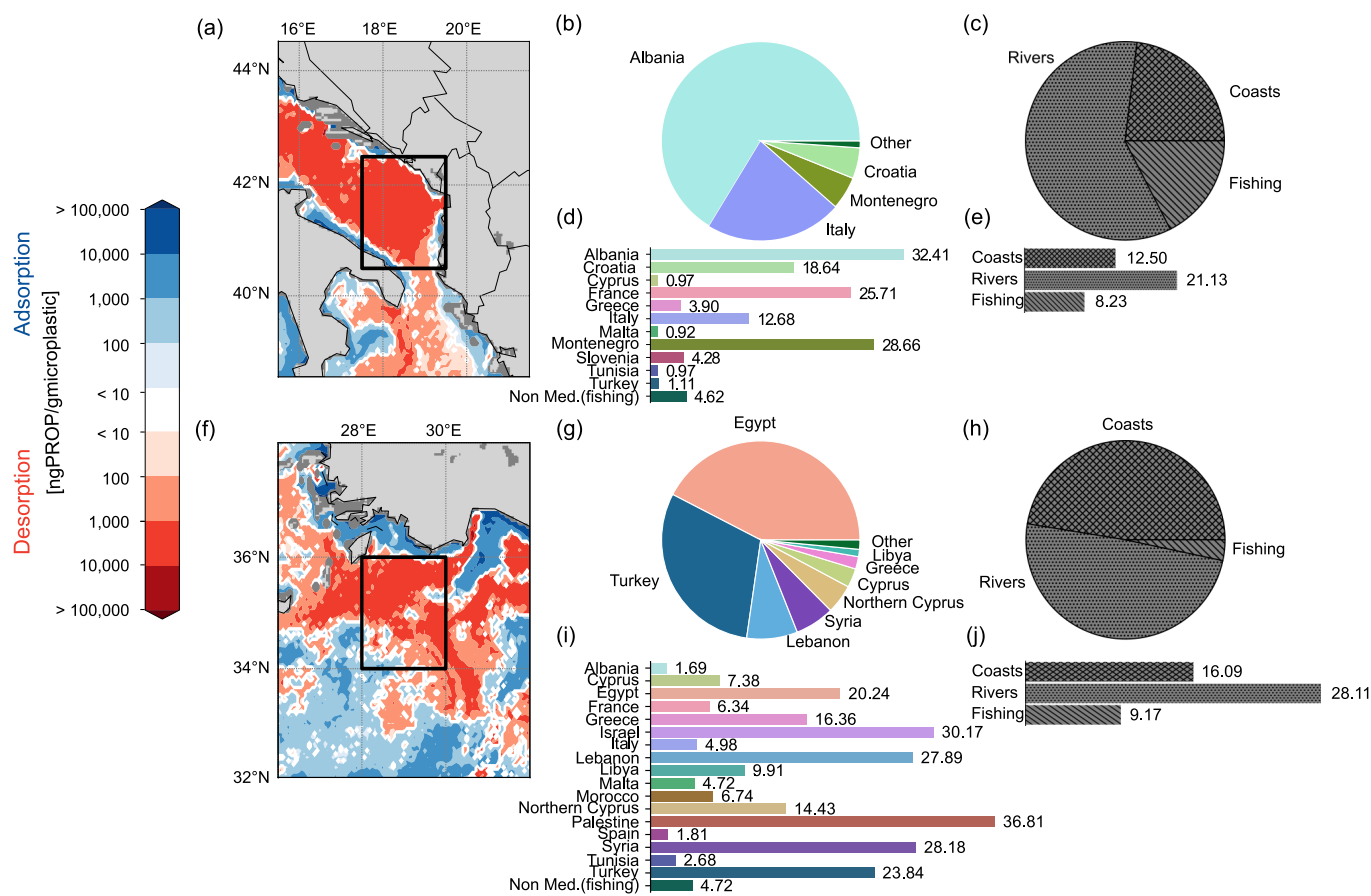}
	\caption{Sourcing of desorbing particles and related PROP apportionment obtained by our coupled model (as in Fig.~\ref{fig:3}(a)) for two exemplificative desorption areas, located in the southern Adriatic Sea (black frame in (a)) and in the northern Levantine Sea (black frame in~(f)); (b) Apportionment of the particles desorbing in the southern Adriatic Sea by country of origin; (c) Source types of desorbing particles; average mass of PROP desorbed by each particle (in ng$_{PROP}$/g$_{particle}$) by country of origin (d) and source type (e). (g)--(j) As in (b)--(e), for the desorption area in (f).}
	\label{fig:4}
\end{figure}

Lagrangian tracking permits to apportion the microplastic-mediated chemical pollution, resulting from the coupling with Eulerian advection-diffusion of PROPs at sea, to specific source types and countries of origin of microplastic particles. The origin of desorbing particles is elaborated for two exemplificative areas, one located in the southern Adriatic ((a)--(e) in Fig.~\ref{fig:4}) and another in between Crete and Cyprus (North-Eastern Mediterranean, (f)--(j)). In the Adriatic area (\ref{fig:4}(a)), the largest number of PROP-releasing particles resulted to originate from the countries of Albania and Italy (\ref{fig:4}(b)), and from rivers as source type (\ref{fig:4}(c)). However, when considering the average mass of PROP desorbed per unit of microplastics' mass, particles from Montenegro, France, and Croatia also figure as significant carriers of pollution (\ref{fig:4}(d)). Among source types, riverine particles released the highest mass of PROP per microplastics into this area (\ref{fig:4}(e)). In the second focal area (\ref{fig:4}(f)), the majority of the desorbing particles originated from countries facing the eastern basin, mostly from Egypt (see the outstanding input from the Nile river and its delta, Fig.~\ref{fig:4}(h) and Fig.~S1), and Turkey. Interestingly, a small, yet significant number of particles comes also from Libya, in the southern Mediterranean Sea (\ref{fig:4}(g)). When sourcing desorbing particles, neither the area nor the population of the country of origin seem to play a significant role there in determining the average mass of PROP desorbed per gram of microplastics, as suggested by the high values attributable to particles released by relatively small countries, such as Lebanon (\ref{fig:4}(h)). Coastal and riverine particles appear to carry and desorb, on average, a higher amount of PROP in the surrounding seawater than fishing-related ones (\ref{fig:4}(j)).

\begin{figure}
	\centering
	\includegraphics[width=180mm]{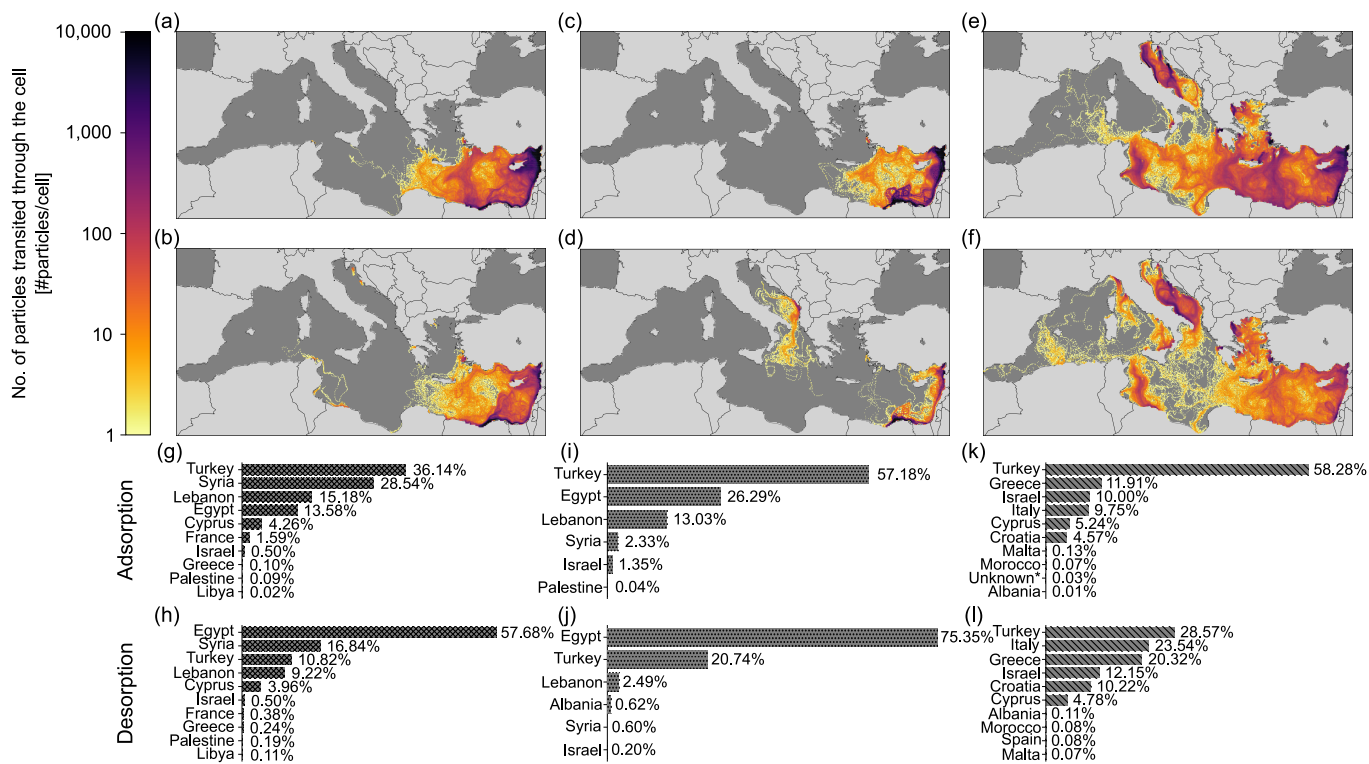}
	\caption{Trajectories (maps) and countries of origin (barplots) of the particles ranking in the top-1\% PROP-adsorbers or -desorbers per source type (see text) according to our coupled model. In the upper maps, the color scale refers to the number of top-particles transited in the reference cell throughout the two years of simulation; the lower barplots show the apportionment of the top-1\% particles by source country. Apportionment by country of origin of the top-1\% ((a), (g)) adsorbing, and top-1\% ((b),(h)) desorbing particles from coasts; ((c), (i)) top-1\% adsorbing, and ((d), (j)) top-1\% desorbing particles from rivers; and ((e), (k)) top-1\% adsorbing, and top-1\% desorbing particles from fishing. The symbol * refers to particles related to fishing vessels with unknown flag.}
	\label{fig:5}
\end{figure}

Fig.~\ref{fig:5} shows the trajectories (maps in (a)--(f)) and the country of origin (barplots in (g)--(l)) of the most chemically active particles per source type, i.e.~particles whose total PROP mass, either adsorbed or desorbed during their whole journey at sea, ranks in the top-1\% when considering the two processes and source types separately. Most of these top-exchanging particles display trajectories confined to the easternmost part of the Mediterranean Sea, independent of their source type (\ref{fig:5}(a)--(f)). Notably, top-exchanging particles released from fishing grounds (\ref{fig:5}(e),(f)) are found also in the southern and central parts of the Mediterranean basin, with some trajectories reaching the Alboran Sea. Top-adsorbing particles (\ref{fig:5}(a),(c),(e)) are more frequently found in coastal waters than top-desorbing ones (\ref{fig:5}(b),(d),(f)). Top-desorbing particles originating from rivers and fishing activities (\ref{fig:5}(d),(f)) followed trajectories that remarkably differ from those of their top-adsorbing counterparts (\ref{fig:5}(c),(e)), as the former reach areas further west. The countries of origin of the top-exchanging particles (\ref{fig:5}(g)--(l)) mostly face the eastern Mediterranean Sea, with the exception of Albania, Croatia, France, Italy, and Spain. Among the eastern Mediterranean countries, particles released from Egypt are the most frequent among the top-desorbing coastal and riverine particles (\ref{fig:5}(h),(j)), while Turkey is the prevalent origin of top-adsorbing particles, independently of their source type (\ref{fig:5}(g),(i),(k)). Western countries account for less than 10\% of the most active particles (\ref{fig:5}(g)--(l)), with the exception of Italy, from where more than 23\% of the top-desorbing fishing-related particles is released (\ref{fig:5}(j)).

\section{Discussion}
\label{sec:4}
We presented a first modelling framework designed to account for the chemical exchanges between microplastic particles and PROPs \citep{Guerrini2021} in marine environments by coupling the dynamics of (Lagrangian) particles and (Eulerian) PROPs within the sea surface layer. The novel axis that the problem at hand required us to explore is that plastic particles both (i) act as passive tracers responding to environmental conditions, and (ii) alter the pollution status of the sea via bidirectional chemical interactions. From a technical perspective, we solved a subproblem (SP3) by turning passive Lagrangian particles into active agents that are able to alter the spatial distribution of PROPs (SP2), while enriching the information brought by particle movement (SP1) by adding information at the particle scale derived from such two-way interactions. Although we are dealing with abiotic materials, some links exist between our approach and Lagrangian biophysical models, where the perception of relevant environmental variables can directly influence the trajectory of particles (see for example \citealp{DiFranco2012,Lett2020}, about the modelling of fish larvae, or of fish otoliths, \citealp{Zenimoto2011,Calo2018}).

In our model, plastic particles are assumed to be unpolluted at release: consequently, they start taking up the focal PROP simulated here from the surrounding (polluted) environment. Interestingly, maps of the average particle-bound PROP concentration (Fig~\ref{fig:2}(b)) do not perfectly match those of average distribution of such PROP in seawater (Fig.~\ref{fig:2}(a)). Particle history (in terms of source type and location, and of journey at sea) clearly influences the particle-bound concentration of the target PROP, as this depends on the chemical gradients experienced by the particle itself during transport, as well as on the consequent PROP exchanges. When quantifying these exchanges, PROP adsorption on particles prevails throughout the Mediterranean basin (Fig.~\ref{fig:3}(a)), as expected because of the initial conditions at release. Contrasting Fig.~\ref{fig:3}(a) to Fig.~S5, we observe that the target PROP tends to partition on particles mostly (but not exclusively) close to source locations and especially along the coasts, from where 80\% of the simulated particles is released (coastal and riverine sources). After release, transport of particles by surface currents contributes to exporting the PROP from coastal areas to high seas. This is the case, for example, of the desorption area in the Levantine basin (Fig.~\ref{fig:3}(a) and Fig.~\ref{fig:4}(f)) resulting from our simulations, that can be attributed to particles released from the densely populated areas of the Nile Delta and the Nile River itself (Fig.~\ref{fig:4}(g)). This example is notable, not only because the number of particles and the amount of PROP released there are the highest in the whole basin (Fig.~S1), but also because our simulations identify such area as highly polluted (Fig.~\ref{fig:2}(a)). South-eastern Mediterranean coastal waters also play a pivotal role for top-exchanging particles, with most of the top-1\% adsorbing particles from coastal and riverine sources passing through them (Fig.~\ref{fig:5}(a),(c)), and Egypt being the origin of >50\% of the top-desorbing particles from these source types (Fig.~\ref{fig:5}(g),(i)). Long-range export of the simulated PROP is also visible in the second focal area examined here (Fig.~\ref{fig:4}(a)): most particles come from the northern Adriatic, such as from the Italian river Po, and countries like Croatia and Montenegro (Fig.~\ref{fig:4}(b)), also carrying significant amounts of PROP (Fig.~\ref{fig:4}(d)). Other relevant particle-mediated fluxes of PROP from coasts to high seas are commented in SM, Section~S5. 

Our simulations suggest that particles could convey PROPs not only from coastal waters to the open sea (reinforcing findings by \citealp{DeFrond2019}), but also across entire sub-basins at regional scales (Fig.~\ref{fig:5}; see also \citealp{Mari2020,Guerrini2021} for Mediterranean connectivity). In particular, the likelihood of microplastics to serve as long-range vectors of PROPs seems quite dependent on their source of release. We found that riverine particles are the most significant vectors of our PROP of reference in the Mediterranean basin, largely contributing to the vastest desorption areas (Fig.~\ref{fig:3}(c)) on top of conveying the highest amounts of the target PROP per gram of microplastics to some of them (Fig.~\ref{fig:4}(e),(l)). Since all particles were released clean, these results can be entirely ascribed to adsorption of PROP occurring within the sea, and not to processes occurring before release (e.g.~industrial effluents within the river catchment). 

When looking at the apportionment of pollution by particles' geographic origin (Fig.~\ref{fig:4}), no clear one-to-one relationship between particle counts and PROP desorption emerged. Some countries stand out in terms of the amount of PROP conveyed by particles originating from them, despite contributing incomparably less than others in terms of particle counts: each country, no matter how small, is accountable for some level of microplastic-related pollution in the Mediterranean Sea. On the one hand, in fact, neither the size nor the population of a country solely determines the quantity of plastics released, because plastic waste mismanagement also enters into play. Translated into real terms, even a slight improvement in a country's waste management effectiveness, or a higher efficiency in blocking/preventing microplastics to enter the sea, can dramatically cut the effects of microplastic and PROP marine pollution in the long run. On the other hand, our modelling results suggest that curbing the input of microplastic particles might not be enough to tackle the ecotoxicological risk posed even by few, yet highly polluted microplastics. 

To effectively fight the environmental threat posed by (micro)plastic pollution to the Mediterranean Sea and the global oceans we ought not to forget about marine chemical pollution. Microplastics themselves can be considered as a persistent organic pollutant, according to the Stockholm Convention definition \citep{Galloway2017}: microplastics may be removed from the sea surface through sinking, but it does not disappear from the ocean, as highlighted by the astonishing amounts of litter and microplastics found in the water column and benthic environments \citep{VanCauwenberghe2013,Choy2019,Kaandorp2020,Kane2020,Kvale2020}, and on shorelines \citep{Onink2021}. Similarly, PROPs are also conveyed to the sea depths by partitioning on suspended sediments (see e.g \citealp{UNEP1992} and \citealp{Merhaby2019}) and on sinking plastic particles (Fig.~\S6). Bottom sediments may be accumulating several types of pollutants, with unforeseeable consequences for benthic ecosystems and, consequently, for the whole marine environment \citep{Kane2020}. Future modelling efforts in this field should account for pollutant fluxes to the seafloor and, in general, for their distribution along the water column (as suggested by \citealp{Jalon-Rojas2019,Soto-Navarro2020} and \citealp{Fuente2021} for microplastics), to extend risk assessment procedures from microplastics on the ocean surface \citep{Everaert2020} to PROPs and their related vertical distributions. 

The major limitation faced by novel, holistic modelling approaches like ours resides in lack of data. Very little is known yet about the magnitude of PROP influxes to the Mediterranean Sea: this knowledge gap heavily influences quantitative assessments of modeled concentrations of PROPs, both in seawater and particle-bound. The latter are determined by kinetics and related parameters that can be studied and retrieved from laboratory evidence (see~SM, Sections~S1 and~S2), and that could be used to inform seawater-plastic partitioning of a focal PROP in our model. Our approach would be potentially capable of more accurate projections if forced with actual influxes of PROPs, in terms of both quantities and source types. For instance, industrial activities \citep{Tobiszewski2012} on land, as well as commercial shipping \citep{UN/MAP2017,UNCTAD2020} and drilling rigs \citep{Cordes2016} offshore, should be encompassed as PROP sources. Furthermore, atmospheric deposition should also be accounted for as input of marine pollutants in the Mediterranean Sea \citep{Castro-Jimenez2014}. 

Finally, the interactions between microplastics and PROPs are strongly affected by polymer-chemical pairing \citep{Rochman2013,OConnor2016,Wang2018,Tourinho2019,Menendez-Pedriza2020}. To exemplify the application of our approach, here we focused on polyethylene particles and used phenanthrene as target PROP (SM, Section~S2) to take advantage of their high affinity and conservatively overestimate particle-bound PROP concentrations \citep{Teuten2007,Amelia2021}. Further modelling efforts could account for the variety of polymers among microplastics \citep{Andrady2017,Rochman2019} and for the simultaneous presence of different PROPs at sea \citep[e.g.~]{Albaiges2005}. 

\section{Conclusions}
\label{5}
Our results highlight that there is no matter of priority in deciding whether to counteract plastic pollution or chemical pollution, because they are inextricably coupled. We believe that both must be dealt with using holistic approaches, as they inherently are two sides of the same coin, not only in terms of their intertwined environmental dynamics, but also for the societal issues they pose, whereby a convenient, yet unsustainable business-as-usual production scenario has to be contrasted with the prospect of causing ecological harm. To prevent unacceptable environmental change, the identification of "safe operating spaces" is a vital, albeit cumbersome task \citep{Rockstrom2009,MacLeod2014,Galloway2016,Villarrubia-Gomez2018,Rockstrom2021}. Healthier seas and oceans for the future generations \citep{UN2020} need the implementation of comprehensive approaches, not only to identify and protect the mechanisms that interconnect marine ecosystems, but also to understand the role played by the processes that are disrupting them. Quantitative methods, such as data-driven numerical modelling like ours, can shed some light on phenomena that are hard to investigate on the field. Approaches that couple physical risks for biota caused by microplastic pollution to chemical hazards related to PROPs, subject to biomagnification effects, are crucial to inform and prioritize actions on key pollution sources, a pivotal task for marine conservation.

\small{
\subsection*{Conflict of Interest Statement}
The authors declare that the research was conducted in the absence of any commercial or financial relationships that could be construed as a potential conflict of interest.

\subsection*{Author Contributions}
All the authors substantively contributed to design and perform the research, other than to writing the paper. Numerical simulations were performed by FG. All authors approved the final version of the paper.

\subsection*{Funding}
This study received funding from the H2020 project “ECOPOTENTIAL: Improving future ecosystem benefits through Earth observations” (grant agreement No. 641762, \href{http://www.ecopotential-project.eu}{http://www.ecopotential-project.eu}).
}

\bibliographystyle{dcu}
\bibliography{biblio}

\end{document}

\maketitle
\section{Characterization of the sources of microplastic pollution in SP1
}
\label{S1}
The Lagrangian tracking algorithm developed to deal with SP1 required to define relevant sources of microplastic pollution. Around 100 million particles per year were allocated among three different source types (coasts, rivers and fisheries) according to the 50:30:20 coasts-to-rivers-to-ships ratio proposed by \cite{Liubartseva2018}. Following the Lagrangian framework developed in \cite{Guerrini2021}, we adopted data-informed, source-specific proxy indicators to modulate the temporal and geographical release patterns from each particle source, defined as follows.

\begin{figure}[h]
	\centering
	\includegraphics[width=180mm]{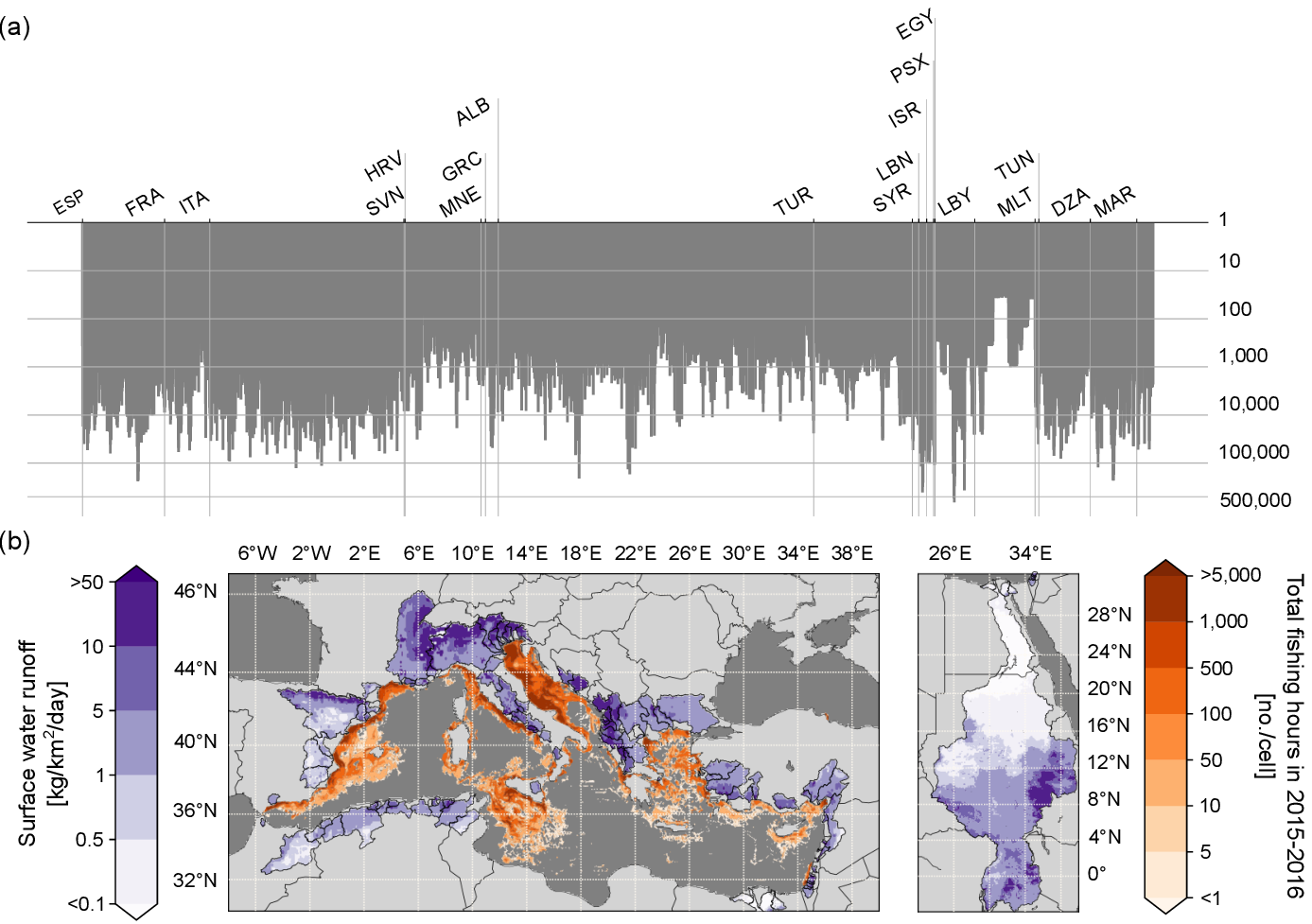}
	\caption{Proxy variables used to characterize particle sources in SP1. (a) Population in the coastal cells (5~km~x~5~km) of the CIESIN dataset, plotted following the Mediterranean coast clockwise, starting from the Strait of Gibraltar. (b) Average surface water runoff in the 100 river basins selected (purple color scale) and total fishing hours in 2015--2016 (orange color scale).
	}
	\label{fig:S1}
\end{figure}

The mismanaged waste produced along the Mediterranean coastlines was selected as a proxy indicator of coastal microplastic input. Coastal sources are assumed to emit a number of particles proportional to the product between two factors: the country-specific per capita estimate of mismanaged plastic waste generation \citep[from][]{Jambeck2015} and the population inhabiting in a 5 km-wide strip along the coast (from the Gridded Population of the World--Population Count 2015 dataset, in Fig.~\ref{fig:S1}(a); \citealp{CIESIN2018}). Overall, a total of 50 million particles per year were released by coastal sources.

Riverine sources allow us to account for the contribution of inland locations to marine plastic pollution. As the hydrological regime strongly affects the presence of debris in river waters \citep{Schmidt2017,VanEmmerik2018}, our riverine proxy indicator was designed to account also for seasonal patterns in surface water runoff. The runoff was retrieved from the gridded FLDAS global dataset \citep{McNally2018}, considering the 2015--2016 period at a monthly scale over all the river basins that discharge directly into the Mediterranean Sea. The riverine proxy indicator was therefore obtained by summing over each basin the cell-by-cell product of (i) monthly surface runoff, (ii) inhabiting population, and (iii) the country-specific estimated fraction of mismanaged plastic waste per capita, using the same gridding as the CIESIN population data (the finest among the three datasets involved in the operation). In other words, we particularized for river basins the procedure explained above for coastal input, while also accounting for the hydrological regime modulating the temporal patterns of particle release. To limit our attention to the most significant rivers in terms of plastic input to the Mediterranean Sea, we averaged the indicator over the study period and selected the top-100 rivers as particle sources for our model (see Fig.~\ref{fig:S1}(b), in purple scale). Overall, these sources released 30 million particles per year, apportioned to each basin and month of simulation proportionally to the proxy indicator detailed above.

Microplastic input from ships focused on fishing activities and was simulated by tracking 20 million particles per year proportionally to an estimate of daily fishing effort at offshore locations retrieved from the Global Fishing Watch dataset (\href{https://globalfishingwatch.org/}{https://globalfishingwatch.org/}; \citealp{Kroodsma2018}) for the years 2015--2016, as in Fig.~\ref{fig:S1}(b) (orange scale). Data processing was required to select fishing data in the Mediterranean Sea, and to remove data regarding fixed fishing gear (like longlines, pots, and traps), as we assume that gear loss happens mostly during handling. 

Additional details and relevant mapping for all plastic sources can be found in \cite{Guerrini2021}.

\section{Model set-up and parameterization for SP2}
\label{S2}

The Eulerian module of the model tackling SP2 had to be run with reference to a marine PROP of reference. We selected the polycyclic aromatic hydrocarbon (PAH) phenanthrene ($C_{14}H_{10}$), because it  has been frequently detected in seawater samples all over the Mediterranean \citep{Berrojalbiz2011} and, most relevantly, adsorbed onto marine plastics \citep{Huffer2016,Wang2018,Leon2019}.
With reference to Eq.~1 (main text, Section~2), the velocity vector $\textbf{v}$ was obtained from the same surface velocities used in SP1 \citep{Simoncelli2014}. The diffusivity coefficient $D$ of phenanthrene in seawater was calculated with the equation proposed by \cite{Schwarzenbach2002}, i.e.
\begin{equation}
\label{eq:S1}
D_{iw}[cm^2 s^{-1}] =\frac{13.26 \cdot 10^{-5}}{\eta^{1.14} \overline{V_{i}^{0.589}}} \, ,
\end{equation}
where $\eta$ is the solution viscosity (in centipoise, 10$^{-2}$g cm$^{-1}$ s$^{-1}$) at the temperature of interest, and $V_i$ is the molar volume of the compound of interest (cm$^3$ mol$^{-1}$), which for phenanthrene is 151.10 cm$^3$ mol$^{-1}$; \citealp{Lide2000}). Diffusivity indeed depends on temperature as well, which may in principle considerably vary because of the Mediterranean-wide spatial domain and the multi-annual span of our simulations. Nonetheless, at temperature values in the range registered for the Mediterranean Sea, the diffusivity values calculated with Eq.~\ref{eq:S1} for several seawater viscosity values \citep[from][]{Nayar2016} were of the same order of magnitude as those obtained using the average surface annual temperature over the whole basin of around 20$^\circ C$ \citep{Shaltout2014}. We then set the obtained $D_{iw}$ constant over the whole oceanographic domain and the entire simulation period. 

In our simulations, input locations of the PROP (where the source term of Eq.~1 in the main text is active) coincide with those for plastic particle input implemented in SP1 (coasts, rivers, and fisheries). To reconcile the two types of inputs, we imposed daily releases of a calibrated amount of pollutant from each source type and location, according to the same apportionment methodology used for Lagrangian particles (see Section~\ref{S1} above). More precisely, we ran several 10-year-long simulations (2005--2014) of SP2 each of which had an arbitrary, yet constant influx of PROP distributed among inputs as mentioned above, and starting from a clean sea surface. Because plastic particle release was modulated over space and time through indicators referring to two years 2015--2016, this is possibly too short of a period to avoid purely transient dynamics of our Eulerian module applied to a clean sea. In fact, and in contrast with the methodology used in SP1, here we chose not to implement any decay of pollutant concentration over time. This is a conservative assumption: in fact, biogeochemical processes such as uptake by phytoplankton and adsorption on sediments are known to mediate vertical fluxes of persistent organic pollutants \citep{Dachs2002,Wu2016}. Following the procedure proposed by \cite{Guerrini2021}, we fed the model with a synthetic time series of inputs where the two-year-long input signals were repeated five times, so as to obtain inputs for a 10-year-long warm-up period. Among the arbitrary fixed values of total PROP influx mentioned above, we then selected the one that, at the end of the simulation, yielded the most realistic PROP concentrations in terms of both numerical values and surface distribution when compared with the (still few) available sampling data (see \citealp{Guerrini2021}, Supplementary Materials, for visual reference). This calibration procedure has been used to circumvent the lack of suitable data about PROP release in the Mediterranean Sea, as mentioned in the main text. 

\section{Model setup and parameterization for SP3}
\label{S3}

\subsection{Particle characteristics}

To deal with SP3, particles needed to be characterized for their polymer type, size and shape, aspects that in SP1 have been neglected (see again Section~SP1 in the main text). In particular, we selected polyethylene (PE = 0.92 $\frac{kg}{l}$; \citealp{WYPYCH2012172}) as polymer type for the microplastic particles of the coupled simulation of SP3. Polyethylene is, in fact, one of the most prevalent polymers in floating marine microplastic litter in the Mediterranean Sea, according to surface sampling campaign data \citep{ErniCassola2019}, while at the same time being particularly relevant in terms of its capacity to sorb organic pollutants \citep{Wang2018}. It has also been observed that the average plastic particle size in Mediterranean samplings is around 1~mm \citep{RuizOrejon2016, Suaria2016, Guven2017, Ruiz-Orejon2018, Zeri2018, Palatinus2019}. 

An important aspect of SP3 is that of accounting for a realistic number of microplastic particles. Although the number of particles released in SP1 was quite high, it is still far from the estimates informed by field counts of microplastic fragment concentrations, which sometimes reach figures in the order of millions per square kilometer \citep{IUCN2020}. As a workaround to achieve plausibly similar particle densities without deploying massive computational power, we estimated a particle-to-microplastics conversion factor for the Mediterranean Sea by comparing the Lagrangian particle counts from SP1 with the densities found at different sampling locations during campaigns carried out in 2015--2016 \cite{Gundogdu2017, Guven2017, Constant2018, Schmidt2018, Zeri2018, DeHaan2019, Palatinus2019}. As a result, we found that each Lagrangian particle could represent about 30,000 microplastic fragments, in a similar way to the super-individual approach used in ecological modelling \citep{Scheffer1995}. Therefore, our prototypical (and theoretical) particle was defined as a sphere with a 1~mm diameter, weighting (and thus exchanging PROP) as 30,000 polyethylene spheres of that size.

From a technical perspective, chemical exchanges between the PROP adsorbed on particles and its presence in waters of the relevant oceanographic cell were assumed to occur at the particle surface, implying homogeneous PROP concentration over particles. Accordingly, PROP uptake by particles has been referred to as "adsorption", and we neglected any intra-particle diffusion driven by PROP gradients within the plastic particle. This assumption is supported by experimental findings on plastic-water exchange for PAHs, whose limiting process is external mass transfer rather than intra-particle diffusion \citep{Endo2013,Li2020}. Furthermore, external mass transfer appears to be especially relevant for compounds with similar or higher weight than phenanthrene \citep{Endo2013,Seidensticker2017}, our target PROP.

\subsection{Plastic-water partition coefficient and desorption rate}
\label{kcoeff}
The two parameters of Eq.~2 (Section~SP2 of the main text) need to be characterized according to the chemical properties of both the selected PROP and the particle polymer type. 

The plastic-water partition coefficient, $K_{P-W}$, is defined as the ratio between the concentration of the pollutant in polyethylene and in seawater at thermodynamic equilibrium. We adjusted the plastic-water partition coefficient provided by \cite{Seidensticker2019} for our target PROP phenanthrene in pure water ($K_{P-W,0}$ = 15,670 L/kg) to account for the average ionic strength of seawater using Setschenow's equation \citep{Setschenow1889}, i.e.
\begin{equation}
\label{eq:S2}
K_{P-W,w} =K_{P-W,0} 10^{[salt]K_s} \, ,
\end{equation}
where the salinity-adjusted value of $K_{P-W,w}$ depends on is the average ionic strength of Mediterranean seawater [salt] and on the so-called Setschenow constant $K_s$. \cite{Lohmann2012a} suggests a typical value of $[salt]$ = 0.5 mol/L (or 30 psu) for seawater, but here we set $[salt]$ = 0.65 mol/L to account for the higher average salinity of Mediterranean waters (about 38 psu, \cite{Brasseur1996}). The Setschenow constant $K_s$ was set at 0.272 for phenanthrene \cite{Karapanagioti2008, Lohmann2012}. We did not account for changes in salinity in either the oceanographic domain nor space, as we pre-verified that those variations had only minor effects on $K_{P-W,w}$; this is especially relevant for phenanthrene, as also shown by empirical evidence \citep{Bakir2014}. We neglected the effects on $K_{P-W,w}$ of other relevant parameters, such as temperature, pressure, and pH \citep{Lohmann2012a}, only because we found those parameters to have only minor effects on the numerical value of $K_{P-W}$ in our study area. Thus, we set $K_{P-W}$ to a value of 23,408 L/kg.

The desorption rate of the target PROP from particles, $k_{des}$, was obtained with a formula introduced by \cite{Endo2013}, i.e.
\begin{equation}
\label{eq:S3}
k_{des} = \frac{D_{w}A_P}{\delta V_P K_{P-W}} \, ,
\end{equation}
where the diffusivity coefficient $D_w$ of phenanthrene in water was set coherently to the one obtained from Eq.~\ref{eq:S1}; the surface ($A_P$) and the volume ($V_P$) of the microplastic particles were set to values specified above, while the thickness of the aqueous boundary layer that surrounds the particle was set to a value of 300 $\mu$m, corresponding to a rather quiescent flow in the ocean environment \citep{Lohmann2012a}.

\subsection{Update of seawater PROP concentration}
In the first step to update the seawater PROP concentration of cell $j$, $C_{W,j}(t)$, one needs to start from the concentrations of PROP on each particle of cell $j$ as calculated with Eq.~2 of the main text. The total mass of PROP exchanged between the cell and the $N$ particles located therein during the interval $\Delta t$, that is $\Delta m_{PROP,j}$, can in fact be calculated as
\begin{equation}
\label{eq:S4}
\Delta m_{PROP,j} = m_P \sum_{i=1}^{N}(C_{i,j}(t+\Delta t)-C_{i,j}(t)) \, , 
\end{equation}
where $C_{i,j}(t+\Delta t)-C_{i,j}(t)$ is the difference between the PROP concentrations on the $i$-th particle found in cell $j$ between times $t+\Delta t$ and $t$, while $m_p$ represents the mass of a single particle, defined above, which is used here to convert the variation recorded in terms of PROP concentration into mass of pollutant transferred. The relative contribution of each particle is then summed up algebraically, duly accounting for its sign, to calculate $\Delta m_{PROP,j}$. This value is finally used to update the PROP concentration in water within the oceanographic cell using the equation
\begin{equation}
\label{eq:S5}
C_{W,j}(t+\Delta t)|_{updated} = \frac{C_{W,j}(t) V_{cell} - \Delta m_{PROP,j}}{V_{cell}} \, ,
\end{equation}
where the water volume of the oceanographic cell ($V_{cell}$) is calculated by considering a parallelepiped of 1/16$^{\circ}$ x 1/16$^{\circ}$ latitude and longitude, and 1~m depth (as in SP1 and SP2; see the related Sections in the main text). $V_{cell}$ is used first to convert the PROP concentration $C_{W,j}(t)$ in cell $j$ obtained from a solution step of SP2 into the total PROP mass in $j$ and then to calculate the new concentration $C_{W,j}(t+\Delta t)|_{updated}$ after the particle-cell interactions. With the update of cell concentrations, SP3 is concluded and each cell undergoes another step of SP2 over [$\frac{\Delta t}{2}$;$\Delta t$, according to the coupling algorithm outlined in Section~2 of the main text.

\section{Processing of the simulation outputs}

The chemical interactions between particles and seawater recorded at a daily resolution over the two-year-long simulation have been analyzed from both an oceanographic and a particle-level perspective. In particular, to identify the spatial patterns in PROP exchanges resulting from the simulation, daily water-particle mass transfers have been first segmented according to the sign of their chemical gradient, so as to distinguish adsorption (from seawater to particles, positive sign) and desorption (from particles to seawater, negative sign). These values have been summed up for each particle, cell-by-cell over the whole oceanographic domain, at every step of the coupled simulation to provide daily maps of adsorption and desorption, respectively. Then, the average specific mass of PROP exchanged per unit of mass of microplastic [ng/g] has been calculated by dividing the daily adsorption/desorption in the cell by the mass of the particles responsible for such exchange of pollutant, i.e. the mass of particles that were adsorbing/releasing PROP. The net particle-mediated flux of PROP in each cell has finally been determined by summing up all the daily signed values of specific adsorption and desorption over the two years of simulation, and the resulting fluxes have been mapped over the oceanographic grid. 

The mapping thus obtained (see Fig.~3 in the main text) allowed the identification of some areas of particular interest, which we used for more in-depth investigations. In particular, the particles that transited through those areas during the simulation were identified, apportioned to their country and source type of origin (see Fig.~4 in the main text). The amount of PROP released by these particles was used to estimate the contributions of the various countries and source types, which were then ranked based on particle count and specific mass of PROP conveyed to the areas of interest through desorption from particles. Additionally, we ranked particles by their total adsorption and desorption of PROP (i.e., by summing up the exchanges mediated by them during their entire journey at sea) separately, identified the top-1\% in adsorption/desorption rankings, and apportioned them to their country and source type of origin. We also extracted from the outputs of SP1 the trajectories at sea of these top-exchanging particles. The aim here was to identify any recurrent pathways and/or particle release locations that could play a role in turning particles into significant vectors of PROP pollution in the Mediterranean Sea.

\section{Apportionment results for other interesting desorption areas of the Mediterranean Sea }
\label{S5}
\begin{figure}[]
	\centering
	\includegraphics[width=180mm]{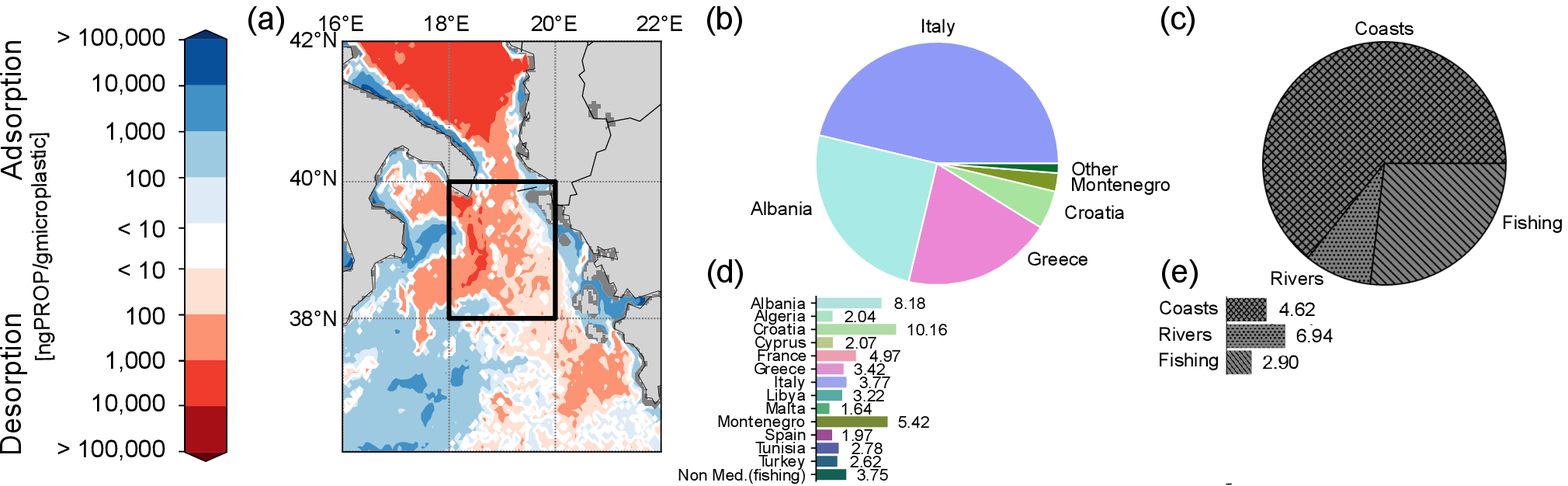}
	\caption{Origin of the desorbing particles in an area with high specific desorption in the Ionian Sea during 2015--2016. (a) Magnification on the Ionian Area (from Fig.~3(a) in the main text). Source apportionment has been performed for particles entering the black plot during the simulated period; (b) Country of origin of desorbing particles; (c) Source type of desorbing particles; (d) Specific PROP desorption per particles' country of origin; (e) Specific PROP desorption per particles' source type.
    }
	\label{fig:S2}
\end{figure}

\begin{figure}[]
	\centering
	\includegraphics[width=180mm]{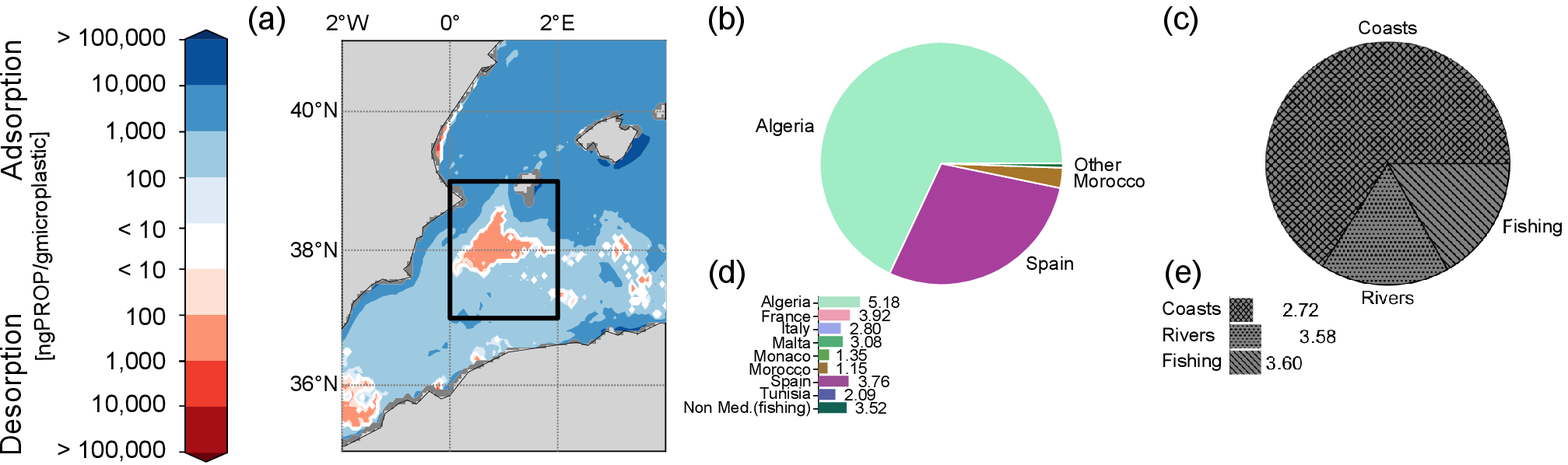}
	\caption{Origin of desorbing particles for an area with high specific desorption in the eastern Alboran Sea during 2015--2016. (a) Magnification on the eastern Alboran Sea (from Fig.~3(a)). Source apportionment has been performed for particles entering the black plot in 2015--2016; (b) Count of desorbing particles within the area per country of origin; (c) Count of desorbing particles within the area per source type; (d) Specific PROP desorption per particles' country of origin; (e) Specific PROP desorption per particles' source type.
    }
	\label{fig:S3}
\end{figure}

\begin{figure}[]
	\centering
	\includegraphics[width=180mm]{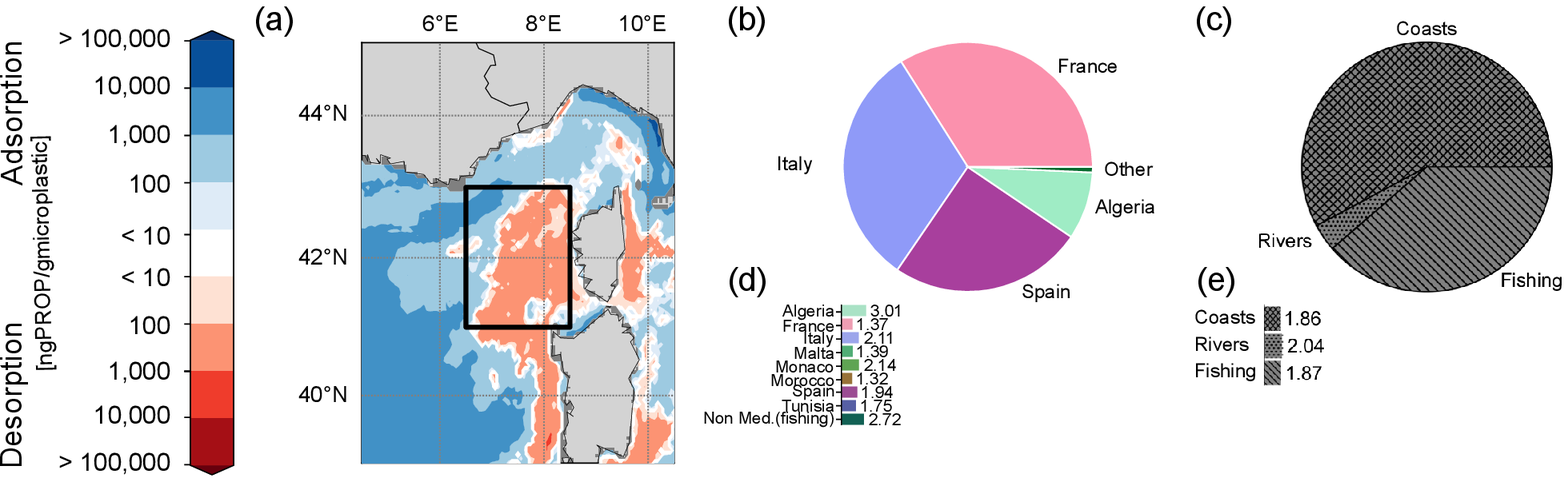}
	\caption{Origin of desorbing particles for an area with high specific desorption in the Sea of Corsica/north-eastern Balearic Sea during 2015--2016. (a) Magnification on the Seas of Corsica and of Sardinia (from Fig.~3(a)). Source apportionment has been performed for particles entering the black plot in 2015--2016; (b) Count of desorbing particles within the area per country of origin; (c) Count of desorbing particles within the area per source type; (d) Specific PROP desorption per particles' country of origin; (e) Specific PROP desorption per particles' source type.
    }
	\label{fig:S4}
\end{figure}

In Figs.~\ref{fig:S2}, \ref{fig:S3}, and~\ref{fig:S4}, we follow the same procedure as in Fig.~4 (main text) to apportion particle-mediated fluxes of PROP in other interesting desorption areas identified in Fig.~3a (main text). 

Fig.~\ref{fig:S2} shows that particles sourced from Italy and from northern Adriatic countries are the main responsible for pollutant desorption in the Ionian Sea in terms of count, and that coasts prevail in terms of source type. This latter result is interesting, as in Fig.~4a, which analyzes an area located just north, riverine particles are the most common (Fig.~4(c)), while in the Ionian Sea rivers account for about 10\% only. This difference can be attributed to the spatial distribution of the rivers included in our model, with many riversheds discharging into the Adriatic Sea (see Fig.~\ref{fig:S1}), and to the surface circulation in the Adriatic sub-basin, which tends to form counterclockwise gyres in the south \citep{Pinardi2000,Millot2005,Liubartseva2018}, possibly retaining particles and preventing them from reaching the Ionian Sea through the Strait of Otranto.

Fig.~\ref{fig:S3} reveals that in the Alboran Sea, south-west of the Balearic islands, desorbing particles are mainly sourced from Algeria, Spain, and France as countries, and from coasts as source type (see Fig.~\ref{fig:S3}(b),(c)). This is also coherent with the results in Fig.~3(a),(b) of the main text, showing that in the coastal waters of these western Mediterranean countries particles are mostly adsorbing PROPs. These PROP-loaded particles are then transported eastwards by surface currents, giving origin to the desorption area analyzed in Fig.~\ref{fig:S3}. These considerations about the west-to-east transfer of PROPs in the Mediterranean basin apply also to the waters on the west side of Corsica (Fig.~\ref{fig:S4}), which however are characterized by less intense desorption than the areas previously analyzed. In fact, a transfer of PROPs from coastal to these high-sea waters can be noticed in Fig.~3(b) (main text), where a small coastal adsorption strip is visible along the coast of Corsica, facing the aforementioned desorption area. Another interesting feature that emerges from Fig.~\ref{fig:S4} is that about half of the particles that release PROPs in this area are sourced from fishing vessels (and not from coasts as suggested by Fig.~3(b) in the main text). Looking again at Fig.~3 (main text), the desorption area next to Corsica is more clearly defined for fishing-related particles (in (d)) than for any other source, and figures as the only fishing-related desorption area located in the western side of the basin. As in this area we found particles from all the countries facing the Alboran Sea, we can reasonably assume that particles released from vessels operating in the Alboran Sea adsorb PROPs during their eastward journey and desorb them in the area under examination. These trajectories are coherent with the general circulation patterns in the western Mediterranean basin, characterized by offshore currents often flowing from the Alboran basin to the Sea of Corsica \citep[see][]{Pinardi2000,Millot2005}. 
\vfill

\begin{figure}[h]
	\centering
	\includegraphics[width=180mm]{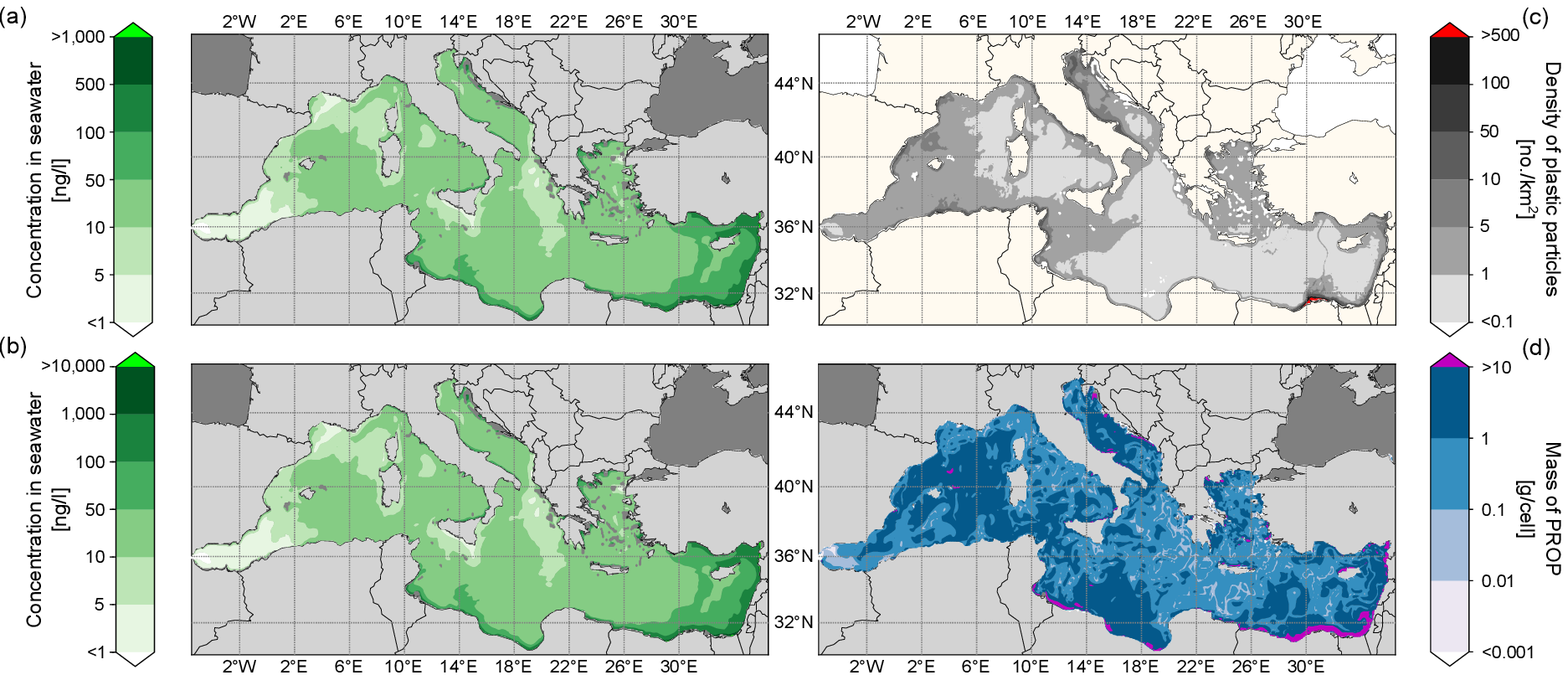}
	\caption{Results of the coupled Lagrangian-Eulerian simulations. (a) Average surface PROP concentrations (sum of PROP mass in seawater and on particles, divided by the volume of oceanographic cells); (b) Average surface PROP concentrations resulting from a particle-free run of the Eulerian advection-diffusion module SP2; (c) Average particle density per square kilometer resulting from the Lagrangian simulations of SP1; (d) Cumulative difference of PROP mass between the purely-Eulerian, particle-free run of SP2 (as in (b)) and the coupled simulations (accounting for both the seawater-dissolved and the particle-bound PROP mass, as in (a)). Despite providing average maps that appear almost identical at a visual inspection, the mass difference between (b) and (a), integrated over the whole spatial simulation domain, amounts to about 45 kg of PROPs. It has to be remembered that SP2 does not involve removal of PROPs from the sea surface, whereas particle-bound PROPs are removed via particle sinking in the coupled Lagrangian-Eulerian simulations. All results refer to the period 2015--2016.
	}
	\label{fig:S5}
\end{figure}

\begin{figure}[h]
	\centering
	\includegraphics[width=90mm]{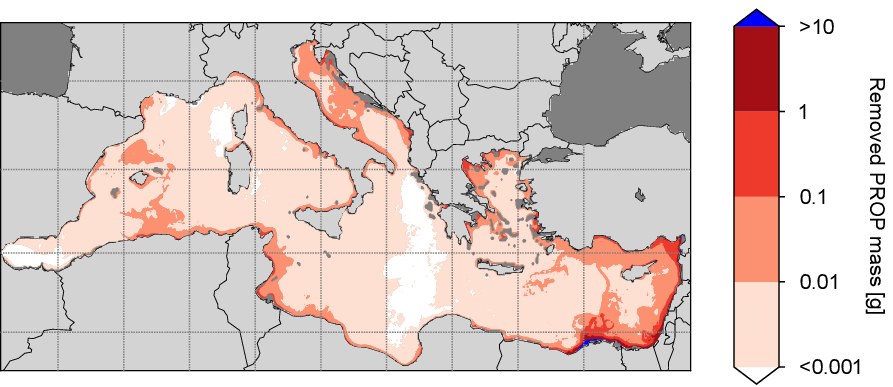}
	\caption{Total amount of particle-bound PROP sunk with particles in 2015--2016.
    }  
	\label{fig:S6}
\end{figure}

\bibliographystyle{dcu}
\bibliography{biblio}